\documentclass[reprint,amsmath,amssymb,aps,pre]{revtex4-1}
\usepackage{url}
\usepackage{amsfonts}
\usepackage{amsmath}
\usepackage{amssymb}
\usepackage{color}
\usepackage{graphicx} 
\usepackage{epstopdf}
\usepackage{bm}
\usepackage{dcolumn}% Align table columns on decimal point
\usepackage{bm}% bold math
\usepackage{hyperref}% add hypertext capabilities
\usepackage{dsfont}
\usepackage[caption=false]{subfig}
\usepackage{float}
\begin{document}

\preprint{APS/123-QED}

\title{Missing Link Identification Using SIS Epidemic Traces}

\author{ Aram Vajdi}
  \email{avajdi@ksu.edu}
\author{Caterina Scoglio}%
 \email{caterina@ksu.edu}
\affiliation{Department of Electrical and Computer Engineering, Kansas State University, Manhattan, Kansas 66506, USA
}
\date{\today}% It is always \today, today,

\begin{abstract}
The study of SIS epidemics on networks has stressed the role of the network topology on the spreading process. However, accurate models of SIS epidemics rely on the complete knowledge of the network topology, which is often not available. This paper tackles the problem of inferring the network topology from observed infection time traces, especially where the network topology is partially known or known with some uncertainty. We propose a Bayesian method to infer the posterior probability of uncertain links in the network, and we derive closed form equations for these probabilities. We also propose a numerical approach based on a Gibbs sampling when the number of uncertain links is large such that using the closed form equations becomes impractical. Numerical results show the capability of the proposed approach to assign high probability to existing links and low probability to non-existing links of the network when the SIS traces are sufficiently long.
\end{abstract}

%\keywords{Complex networks, epidemic spreading, Markov process, Network %inference.}

\maketitle

\section{Introduction}\label{introduction}
\begin{figure}[t]
  \label{FC1} \includegraphics[width=0.75\columnwidth]{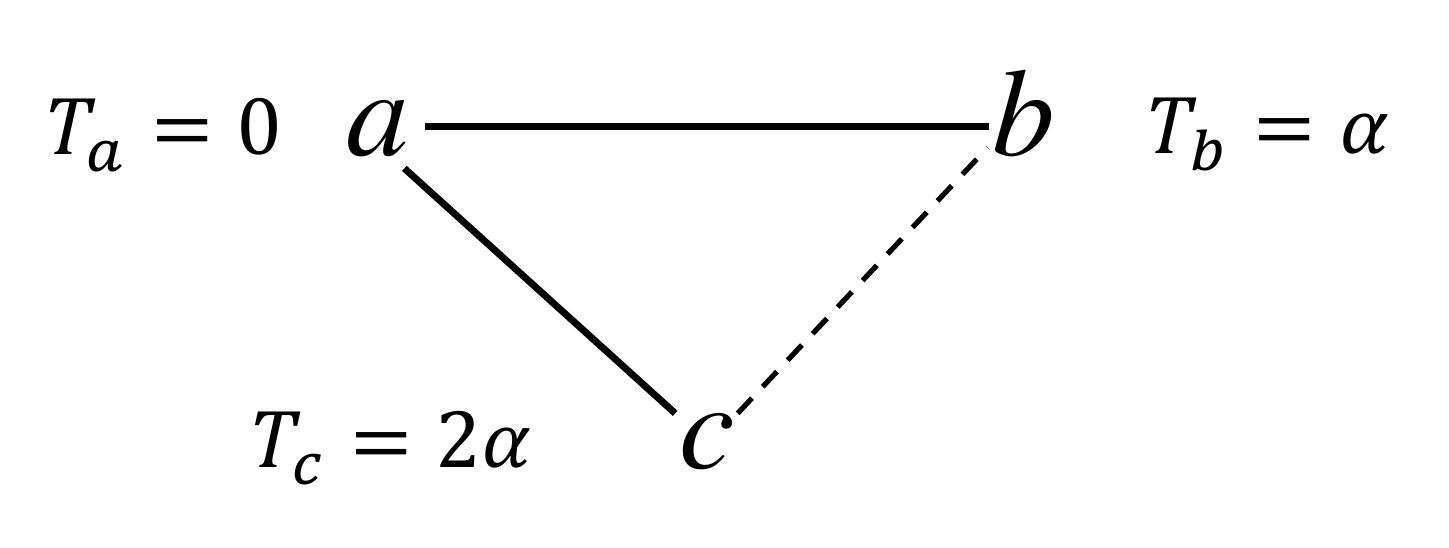} %     
   \caption{In this network, we know the links ab and ac exist, but the link bc is uncertain. Also, we know the expectation time for transmission of infection through any link is $\alpha$. Since the difference between the infection time of nodes b and c equals the expected transmission time $\alpha$, link bc should be present in the network because the infection of node c by node a has a low probability due to the difference between their infection times.}
\end{figure}

To understand and control behavior of an epidemic in a population, researchers have developed various mathematical models to describe spreading processes \cite{anderson1992infectious,keeling2008modeling,ferguson2000more,gupta1989networks,may1988transmission,muller2000ring,eames2002modeling,boccaletti2014structure,naasell2002stochastic,
dorogovtsev2008critical,pastor2002immunization,pastor2015epidemic,van2011n,sahneh2013generalized,
kiss2017mathematics,kephart1991directed,newman2002email}. Particularly, stochastic spreading models over a graph can be used to study the propagation of infectious diseases, dissemination of information and ideas or spread of computer malware. In such a model, the nodes of the graph are the individuals, and the edges represent the possible mean for contagion. Since stochastic models assume contagion happens as a result of random processes, they are particularly useful when the description of contagion at the individual level includes some kind of uncertainty, which can be described using such models. 
Some of the standard and basic epidemic models include SI, SIR and SIS, where  S (Susceptible), I (Infectious) and R (Recovered) denote the states of individuals.
In these models, a susceptible node becomes infectious as a result of interaction with the infectious neighbors in the network. In the SIR model, the infectious nodes recover after a period of time and do not contract or transmit infection again. Conversely, in the SIS model, an infectious node becomes susceptible again. Recently, the effect of the network structure on the epidemic has been an active line of research \cite{pastor2001epidemic,newman2002spread,pastor2001epidemic2,
boguna2003absence,chakrabarti2008epidemic,
goltsev2012localization}. Because the network structure leaves its imprint on the epidemic data, we expect the possibility of recovering some information about the network using the observed epidemic data. This inverse problem can be of particular interest when we have only partial information about the network structure that may render control of spreading impossible. To have an intuitive picture of how to use the epidemic data to make inferences about the network structure, we consider the simple graph in figure \ref{FC1} where the link $bc$ is uncertain. Let's assume we have observed an SI process where the infection times for the nodes $a, b, c$ are $T_{a}=0,\  T_{b}=\alpha$ and $T_{c}=2\alpha$, respectively. If the expected time for infection transmission through any existing link is $\alpha$, with a high probability, we expect that node $c$ has been infected by node $b$, which in turn indicates the existence of the uncertain link $bc$.  

In fact, network structure inference using spreading data has been an active research area in data mining
 \cite{gomez2010inferring,netrapalli2012learning,kamei2012predicting,daneshmand2014estimating,
 choi2015constructing,duong2011modeling,embar2014bayesian,sefer2015convex,
braunstein2016network,li2016learning,altarelli2014bayesian}. One of the earlier works in this field \cite{gomez2010inferring} uses spreading data from a variant of the SI model and tries to reconstruct the whole binary network by an approximation of the maximum likelihood estimator. In a different setting, reference \cite{daneshmand2014estimating} considers another version of the SI model and uses the maximum likelihood estimator to find the best value of parameters in a continuous space. Meanwhile, in \cite{netrapalli2012learning}, the authors assume a discrete time SIR model and use the maximum likelihood estimator to  make inferences about the network structure. Moreover, they address the question about the number of required samples for graph recovery. Specifically, in \cite{netrapalli2012learning,daneshmand2014estimating}, the authors utilize the fact that for the directed networks unlike for the undirected, the global maximum likelihood problem decouples into $N$ local problems, where $N$ is the number of nodes in the network. However, prior information such as the symmetry of links or the range of transmission rates for a link might be available from another source of data quite different from the epidemic data. In such cases, from a practical point of view, it is more plausible to use all the prior information in the inference task. This is particularly useful when we have limited samples of epidemic data, which is often the case in practical application. Although it is possible to include some of the prior information in the maximum likelihood formulation of the problem, we chose the Bayesian method that inherently uses the known facts about the system. Moreover, Bayesian inference, unlike the maximum likelihood method, which is a point estimator, provides a joint posterior distribution for the parameters, which is especially useful in risk assessment.
In this paper, we address the problem of recovering network structure from the traces of continuous time SIS spreading processes. We formulated the problem in the Bayesian framework and inferred the posterior probability for the uncertain links. Here, we assumed a setting where we have the data from a SIS spreading process over a static complex network with some of the links missing. By epidemic data, we mean the history of all nodes' states for a period of time. In other words, we know when the nodes became infected and susceptible again. Moreover, we only considered static underlying networks where the links do not appear or disappear. In this work, we restricted the analysis to a SIS model where the waiting time for the transmission of infection through a link has an exponential distribution with a rate that might be different from that of other links. In addition, we assumed some prior knowledge about the links in the network. In particular, for the exponentially distributed waiting time, prior information about the links can be represented using prior distribution of infection rates.\par
Finally, we want to mention that the network structure inference is not limited to link inference. In existing literature, authors try to find the community structure among the network nodes using epidemic data \cite{barbieri2013cascade,tran2015netcodec}, and in a different setting, the researchers' goal was to identify the source of infection from some observation of epidemic data \cite{altarelli2014bayesian,PhysRevE.90.012801,hu2015network,antulov2015identification,chen2016detecting,PhysRevLett.109.068702}.
% \section{method}
%In this section, after presenting the node %level description of the SIS model, we %derive the likelihood of an SIS trace. %Moreover, we address the Bayesian method of %inference for obtaining the network %structure from the SIS traces. 
\section{SIS model}
In the susceptible-infected-susceptible (SIS) stochastic epidemic model, the individuals are either susceptible or infected. Here, we assumed if a node is infected the probability to stay infected decreases with a constant rate $\delta$. Thus, after the node gets infected at time $t_{0}$, it becomes susceptible again at time $t+t_{0}$ where $t$ is a random variable with the exponential probability density function, $f(t)=\delta \exp(-\delta t)$. 
In this model, the transition from susceptible to infected state is caused by the interaction with an infected neighbor in the network. We assumed, for a susceptible node that has one infected neighbor, the probability to remain susceptible up to time $t$ is $\exp(-\lambda t)$, where $t$ is the duration of contact and $\lambda$ is a constant. In other words, the transmission time for the infection is a random variable that is exponentially distributed with the rate $\lambda$. In practice, a susceptible node may have several infected neighbors trying to infect it. In this case, the infection processes by different neighbors are independent, and the susceptible node gets the infection from the first neighbor that transmits it.
If at $t=0$ node $a$ is susceptible and the infected neighbors set is $\mathcal{N_{I}}$, the probability density function for the infection time, assuming all the  infected neighbors remain infected, is
\begin{equation}
\begin{split}
f(t)=&\sum_{n\in \mathcal{N_{I}}}\ \lambda_{n,a}\exp(-\lambda_{n,a}t)\prod_{n^{\prime}\in\mathcal{N_{I}}-\{n\}}\exp(-\lambda_{n^{\prime},a}t)\\&=\lambda_{0}\exp(-\lambda_{0}t)
\end{split}
\end{equation}
where $\lambda_{0}=  \sum_{n\in \mathcal{N_{I}}}\ \lambda_{n,a}$. In the first line of the equation above, the product term is the probability that the susceptible node $a$ has not been infected up to time instant $t$ by any of  the neighbors $n^{\prime}\in\mathcal{N_{I}}-\{n\}$. The second line in the equation indicates that the minimum of several exponentially distributed random variables has an exponential distribution with a rate equal to the sum of all the  rates for the independent variables. Based on the node level description of the SIS process, clearly the transition of a node from susceptible state to infectious state depends on the state of its neighbor; this implies the dynamics of a node can not be decoupled from those of its neighbors. Instead, the joint state of all the $N$ nodes in the network denoted as $\mathbb{S}=[s_{1},s_{2},\cdots, s_{N}]$, where $s_{i}=1$ ($s_{i}=0$) if $n_{i}$ is infectious (susceptible) , is a continuous-time Markov chain over a space consisting of $2^{N}$ possible network states. 
To derive the likelihood of a network state trace, we first obtain the likelihood of events happening in the network. Assuming at $t=t_{0}$ the network state is $\mathbb{S}(t_{0})$, by an event we mean the first time that a node makes a transition after $t_{0}$. We specify an event by the order pair $e=(n(e),t(e))$ where $n(e)$ is the node that makes the transition and $t(e)$ is the time at which the event happens. Based on the node level description of the SIS process, we can write the probability density function of the event as  
\begin{equation}\label{fe}
\begin{split}
 &f(e\mid \mathbb{S}(t_{0}), \Lambda^{\ast},\delta)=\\&
 \Big(s_{n(e)}(t_{0})\delta_{n(e)}+(1-s_{n(e)}(t_{0}))\sum_{q\in \mathcal{N}^{e}}s_{q}(t_{0})\lambda_{q,n(e)}\Big)\\&\times\exp\Big(-\Delta\sum_{p\in \mathcal{N}^{e}}s_{p}(t_{0})\delta_{p}+(1-s_{p}(t_{0}))\sum_{q\in \mathcal{N}^{e}}s_{q}(t_{0})\lambda_{q,p}\Big)
\\&\times\exp\Big(-\Delta\Big(s_{n(e)}(t_{0})\delta_{n(e)}+\\&\ \ \ \ \ \ \ \ \ \ \ \ \ \ \ \ \ \ \ \ \ \ \ \ (1-s_{n(e)}(t_{0}))\sum_{q\in \mathcal{N}^{e}}s_{q}(t_{0})\lambda_{q,n(e)}\Big)\Big),
\end{split}
\end{equation}
 where $\Delta=t(e)-t_{0}$.
In Eq.(\ref{fe}), $\Lambda^{\ast}$ is the set of interaction rates between the nodes in the network, and $\delta$ is the set of recovery rates. If no link between two of the nodes exist, the interaction rate between them is zero. In this equation, the interaction rate $\lambda_{p,q}$ can be different from $\lambda_{q,p}$, and $\mathcal{N}^{e}$ denotes the set of all the nodes in the network except $n(e)$, $\mathcal{N}^{e}=\mathcal{N}-\{n(e)\}$. 
The second line on the r.h.s of the of Eq.(\ref{fe}) is the probability that none of the nodes in the set $\mathcal{N}^{e}$ makes a transition before $t(e)$, while the first and third lines give the probability density function for the transition time of $n(e)$.
\section{Bayesian inference of missing links}
To use the Bayesian method of inference, we need to calculate the likelihood of observed data conditioned on the parameters. In our problem statement, we assumed we have access to the complete history of nodes' traces for a period of time from $t=0$ to $t=T$. Using the nodes' traces, we can extract all the network states events that occur for that period of time. In other words, the observed data is the sequence of network state events. 
Assuming $\mathcal{C}=\{e_{1},e_{2,\cdots}\}$ is the observed sequence of events ordered by occurrence in time, $t(e_{1})<t(e_{2})<\cdots$, the likelihood of the sequence $\mathcal{C}$ is 
\begin{equation}\label{fc}
f(\mathcal{C}\mid \Lambda^{\ast},\delta)=\prod_{i=1}f(e_{i}\mid \mathbb{S}(t(e_{i-1})), \lambda,\delta)
\end{equation}
where the terms in the product are calculated from Eq.(\ref{fe})
Since we want to make an inference about the interaction rates, in the expression for the likelihood of sequence $\mathcal{C}$, we are only interested in those terms that are a function of $\lambda$. After inserting the density function of events from Eq.(\ref{fe}) into Eq.(\ref{fc}) and absorbing the terms that are a function of $\delta$ into the variable $K(\delta)$, the probability density function of sequence $\mathcal{C}$ simplifies as
\begin{equation}{\label{fc2}}
\begin{split}
f(\mathcal{C}\mid \Lambda^{\ast},\delta)&=K(\delta)\times\exp\Big(-\sum_{q,p\in \mathcal{N}} T_{q,p}\ \lambda_{q,p}\Big)\\&\times\prod_{e_{i}\in\mathcal{C}_{I}}
\sum_{q\in \mathcal{N}}s_{q}(t(e_{i-1}))\ \lambda_{q,n(e_{i})}
\end{split}
\end{equation}
In this expression, $\mathcal{C}_{I}$ refers to the set of all the events $e_{i}$ in the sequence $\mathcal{C}$ that are infecting events, 
$s_{n(e_{i})}(t(e_{i-1}))=0$ and $s_{n(e_{i})}(t(e_{i}))=1$. Here, $s_{q}(t(e_{i-1}))$ is the state of node $q$ just before the event $e_{i}$ happens.
The constant parameter $T_{q,p}$ in Eq.(\ref{fc2}) is the total period of time that node $q$ had the possibility to infect node $p$, in other word
\begin{equation}\nonumber
T_{q,p}=\int_{0}^{T} (1-s_{p}(t))s_{q}(t) \ dt
\end{equation}
The likelihood of sequence $\mathcal{C}$ presented in Eq.(\ref{fc2}) is valid for both directed and undirected networks. When the network is directed $\lambda _{q,p}$ is assumed to be a parameter different from  $\lambda _{p,q}$. Instead, when the network is undirected, $\lambda _{q,p}$ and $\lambda _{p,q}$ refer to the same parameter. Although in deriving the likelihood of the SIS trace we assumed a link between any two nodes $p,q$ with a corresponding rate $\lambda_{p,q}$, the expression in Eq.(\ref{fc2}) is also valid when some of these links are absent. For nonexisting links, we can simply apply $\lambda_{p,q}=0$ and arrive at the correct expression for the liklihood.\par
Now that we have an expression for the likelihood of observed data, we can use Bayes' theorem to find the posterior distribution for the uncertain links. If we use $\Lambda$ to indicate the set of  transmission rates for the uncertain links, Bayes'  theorem gives the joint posterior distribution of the transmission rates as 
\begin{equation}\label{bay1}
\begin{split}
f(\Lambda\mid\mathcal{C})&=\kappa\times\exp\Big(-\sum_{\lambda\in \Lambda} T_{\lambda}\ \lambda\Big)\\&
\times\prod_{e_{i}\in\mathcal{C}_{I}}
\Big(\overline{\beta}_{e_{i}}+\sum_{\lambda\in\Lambda_{e_{i}}}\lambda\Big)\times \mathfrak{F}(\Lambda)
\end{split}
\end{equation}
In this equation, $\kappa$ is a normalization factor, and 
$\mathfrak{F}(\Lambda)$ is the prior distribution of the transmission rates for the uncertain links. Here we have used $\Lambda_{e_{i}}$ to refer to the set of transmission rates for those uncertain links that were active in the event $e_{i}$.  The link $(q,n(e_{i}))$ is active in the infecting event $e_{i}$ if and only if the node $q$ is infectious at the time when the event happens. Moreover, in Eq.(\ref{bay1}),  $\overline{\beta}_{e_{i}}$ is the sum of all the transmission rates for the active links in the event $e_{i}$ except those active links that are uncertain. 
\[
\Lambda_{e_{i}}=\Big\{\lambda_{q,n(e_{i})}\mid\lambda_{q,n(e_{i})}\in\Lambda,s_{q}(t(e_{i-1}))=1\Big\}
\]
\[
\overline{\beta}_{e_{i}}=\sum_{\substack{q\in \mathcal{N}\\\lambda_{q,n(e_{i})}\notin\Lambda}}s_{q}(t(e_{i-1}))\ \lambda_{q,n(e_{i})}
\]
As we mentioned before, if an uncertain link is undirected, both 
$\lambda_{q,p}$ and $\lambda_{p,q}$ refer to the same parameter 
$\lambda \in \Lambda$. In such cases $T_{\lambda}=T_{p,q}+T_{q,p}$. Conversely, when $\lambda\in\Lambda$ refers to a directed link
 $\lambda_{p,q}$, then we have $T_{\lambda}=T_{p,q}$. \par
 The expression in Eq.(\ref{bay1}) provides a joint distribution for the uncertain links. However, we are often interested in some marginal probability distribution such as the distribution of a specific link $\lambda_{0}$. To obtain the posterior distribution $f(\lambda_{0}\mid\mathcal{C})$, we need to integrate the joint distribution over all the other uncertain links. If $\Lambda^{-}=\Lambda-\{\lambda_{0}\}$, we have
\begin{equation}\label{marginal1}
f(\lambda_{0}\mid\mathcal{C})=\int f(\Lambda\mid\mathcal{C})\ d\Lambda^{-}\ .
\end{equation}
Although in some cases this integration is a straightforward task, when we have a large number of uncertain links, the integration might be intractable.
Nevertheless, when the prior distribution in Eq.(\ref{bay1}) is a product of independent factors   
\begin{equation}\label{prior1}
\mathfrak{F}(\Lambda)=\prod_{\lambda\in\
\Lambda}\mathfrak{f}_{\lambda}(\lambda)\ ,
\end{equation}
the integration in  Eq.(\ref{marginal1}) results in a marginal distribution that has a functional form of 
\begin{equation}\label{marginal2}
\begin{split}
f(\lambda_{0}\mid\mathcal{C})&=\kappa_{0}\times\exp\Big(-T_{\lambda_{0}}\lambda_{0}\Big)\ \mathfrak{f}_{\lambda_{0}}(\lambda_{0})\\ &\times \sum_{j=0}^{J}a_{j}\ {\lambda_{0}}^{j}.
\end{split}
\end{equation}
In this equation, $J$ is the number of infecting events that the link $\lambda_{0}$ has been active in, $\kappa_{0}$ is a normalization factor, and the coefficients in the polynomial term can be calculated by performing the integration. However, for a general case when $\lambda_{0}$ is connected to a large number of uncertain links through the factor terms in the joint distribution of Eq. (\ref{bay1}), the integration is not tractable. In such cases, it is possible to apply one of the commonly used numerical methods in Bayesian inference such as the Markov chain Monte Carlo (MCMC) or Belief propagation.
In this article, we use the prior distribution of Eq.(\ref{prior1}) where the independent factors have the functional form as
\begin{equation}\label{pri}
\mathfrak{f}_{\lambda}(\lambda)=\mathrm{P}_{\lambda}\ \delta(\lambda-r_{\lambda})+ (1-\mathrm{P}_{\lambda})\ \delta(\lambda)
\end{equation}
Here, $P_{\lambda}$ is the prior probability for the existence of the link, $r_{\lambda}$ is the transmission rate assuming the link exists, and  $\delta(\lambda)$ is the Dirac delta function.
To find $\mathrm{P}_{\lambda}$, one can use some source of information about the network other than the epidemic traces. For example, when the existence of links between the nodes stochasticly depends on some kind of distance between the nodes,  one can use the known distance to find the prior probability $\mathrm{P}_{\lambda}$. Furthermore, when we do not have any prior information about a link, we can use 
 $\mathrm{P}_{\lambda}=1/2$ as the prior probability. In cases where the integration in Eq.(\ref{marginal1}) is tractable, we can find the posterior probability for existence of the link, $\widehat{\mathrm{P}}_{\lambda}$, from the equation below
\begin{equation}
\frac{\widehat{\mathrm{P}}_{\lambda}}{1-\widehat{\mathrm{P}}_{\lambda}}=\frac{\mathrm{P}_{\lambda}\ \exp\Big(-T_{\lambda}r_{\lambda}\Big)\ \sum_{j=0}^{J}a_{j}\ {r_{\lambda}}^{j}
}{(1-\mathrm{P}_{\lambda})\ a_{0}}\ .
\end{equation}
In practice, for most cases where it is not possible to calculate polynomial coefficients analytically, we can use Gibbs sampling to find the posterior probability $\widehat{\mathrm{P}}_{\lambda}$.
Gibbs sampling requires constructing a Markov chain over the parameters' space that has an equilibrium distribution similar to the joint distribution of the parameters. In our inference problem, the parameters' space is the direct sum of transmission rate space of uncertain links, and the Gibbs sampling constructs a Markov chain $X=(\lambda_{1},\lambda_{2},\cdots,\lambda_{k})$, $\lambda_{i}\in\Lambda$, which has an equilibrium distribution similar to that in Eq.(\ref{bay1}). When we use the prior distribution in  Eq.(\ref{pri}) for the transmission rate $\lambda$, the corresponding component of Markov chain can only assume values of $0$ or $r_{\lambda}$, and using samples from the Markov chain, we can estimate $\widehat{\mathrm{P}}_{\lambda}$
as the fraction of samples with $\lambda=r_{\lambda}$. 
\section{Numerical experiments}
In this section, we report the result of two numerical experiments on synthetic data. In both experiments, we assume if an uncertain link exists, it is undirected, and the infection rate is known . In the  first inference experiment, we do not use any other prior information while in the second experiment, we use a prior probability for the uncertain links which depends on the distance of the nodes.
\subsection{Experiment using uninformative prior}\label{exp1}
\begin{figure}[t]
	\subfloat[]{
		\includegraphics[width=1\columnwidth]{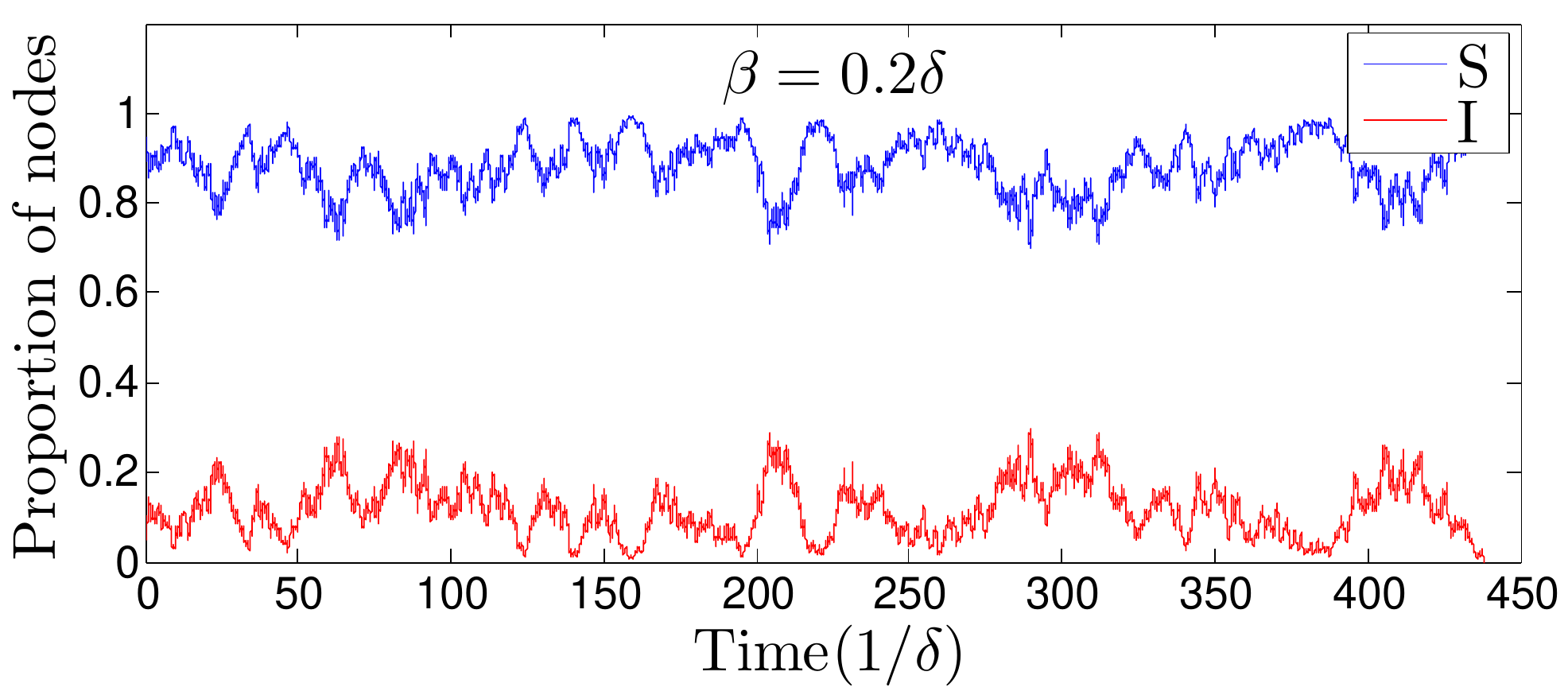}
\label{atr2}} \\
	\subfloat[]{ 
	
		\includegraphics[width=1\columnwidth]{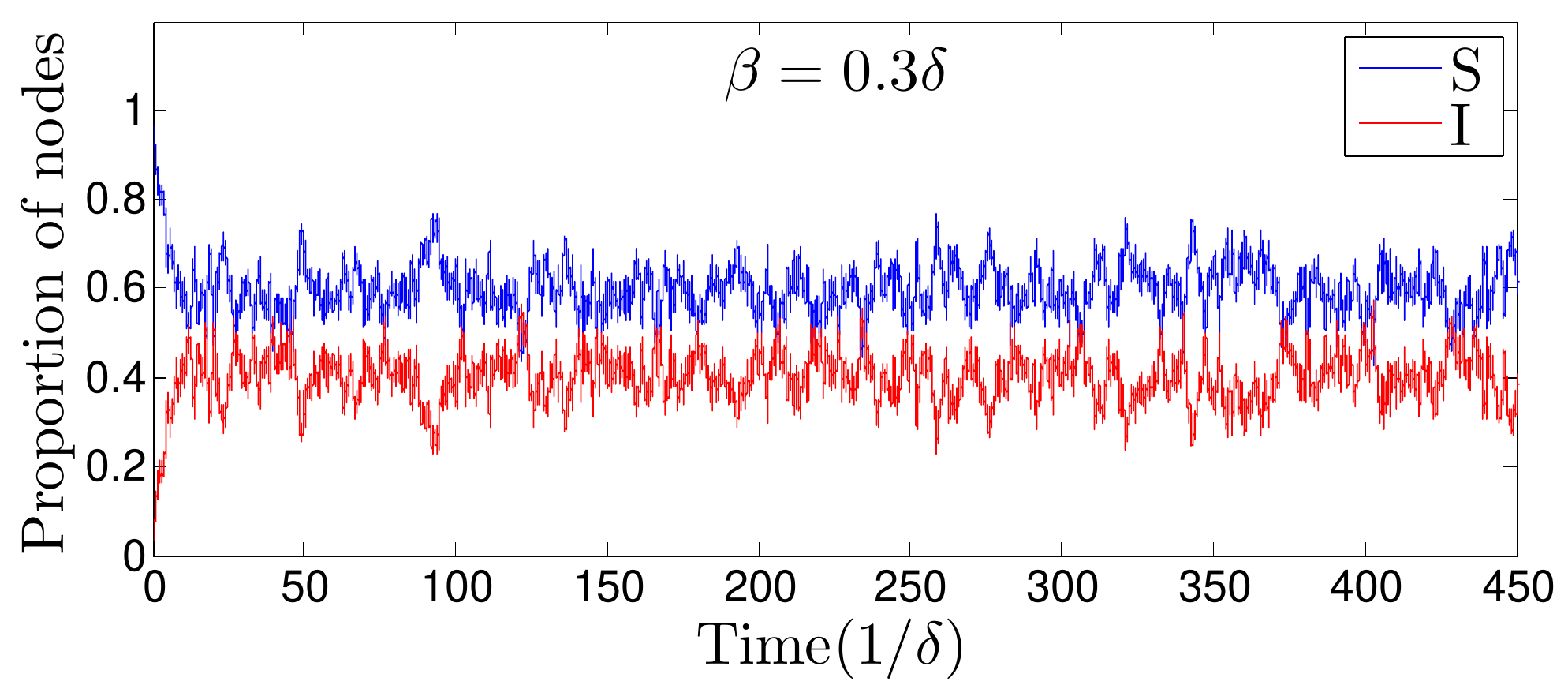}
\label{atr3}}%
	
	\caption{ Proportion of infected and susceptible nodes in the network through time for two different SIS traces used in the experiment in section \ref{exp1}.   } 
	\label{atr}%
\end{figure}

In this experiment, we first generated a random network of $200$ nodes such that any possible edge between the nodes was included in the network with probability $0.03$. This resulted in a random network with $599$ edges and average node degree of $6$. After generating the network, we simulated the SIS epidemic scenarios over the network using an available simulator \cite{sahneh2016gemfsim,GEMFTool}. We assumed the infection starts from $10$ initially infected nodes and spreads through the network links at the same transmission rate. We generated two different SIS traces with two different values for the transmission rates. In the first simulation, we assumed the transmission rate is $\beta=0.2 \delta$
where $\delta$ is the recovery rate for the nodes after they get infected. In the second simulation, we used $\beta=0.3 \delta$ which leads to a larger infected population in the network. Figure \ref{atr} shows the population of infected individuals in the network through time for the two SIS spreading simulations. For $\beta=0.3\delta$, around $40$ percent of nodes in the network are infected at each time instant while for $\beta=0.2 \delta$, this number is smaller.
In this experiment, we assume a set of uncertain links, and we try to find the posterior probability for the uncertain links using any of the SIS epidemic traces.   
In effect, an uncertain link can be an actual link in the network or a link that is not present in the network, and our goal is to determine the probability that it exists. Hence, for the experiment, we assumed a set of uncertain links such that half of them were randomly chosen among the actual network links with the rest being among the non-existing links. For the uncertain link set, we used a prior distribution of the form of Eq.(\ref{prior1})  with 
\[\mathfrak{f}_{\lambda}(\lambda)=\frac{1}{2}\ \delta(\lambda-\beta)+ \frac{1}{2}\ \delta(\lambda).
\]
This prior distribution does not favor the presence or absence of an uncertain link.  However, it limits the transmission rate of a link to a known value of $\beta$. After performing the Gibbs sampling, we obtained the posterior probability for each uncertain link. To show the overall result, we calculated an average error for the uncertain link set  defined as 
\begin{equation}\label{er}
\overline{e}=\frac{\sum_{i=1}^{n_{l}}|\gamma_{i}-\widehat{\mathrm{P}}_{i}|}{n_{ul}}.
\end{equation}

\begin{figure}[t]
  \centering
    \includegraphics[width=1\columnwidth]{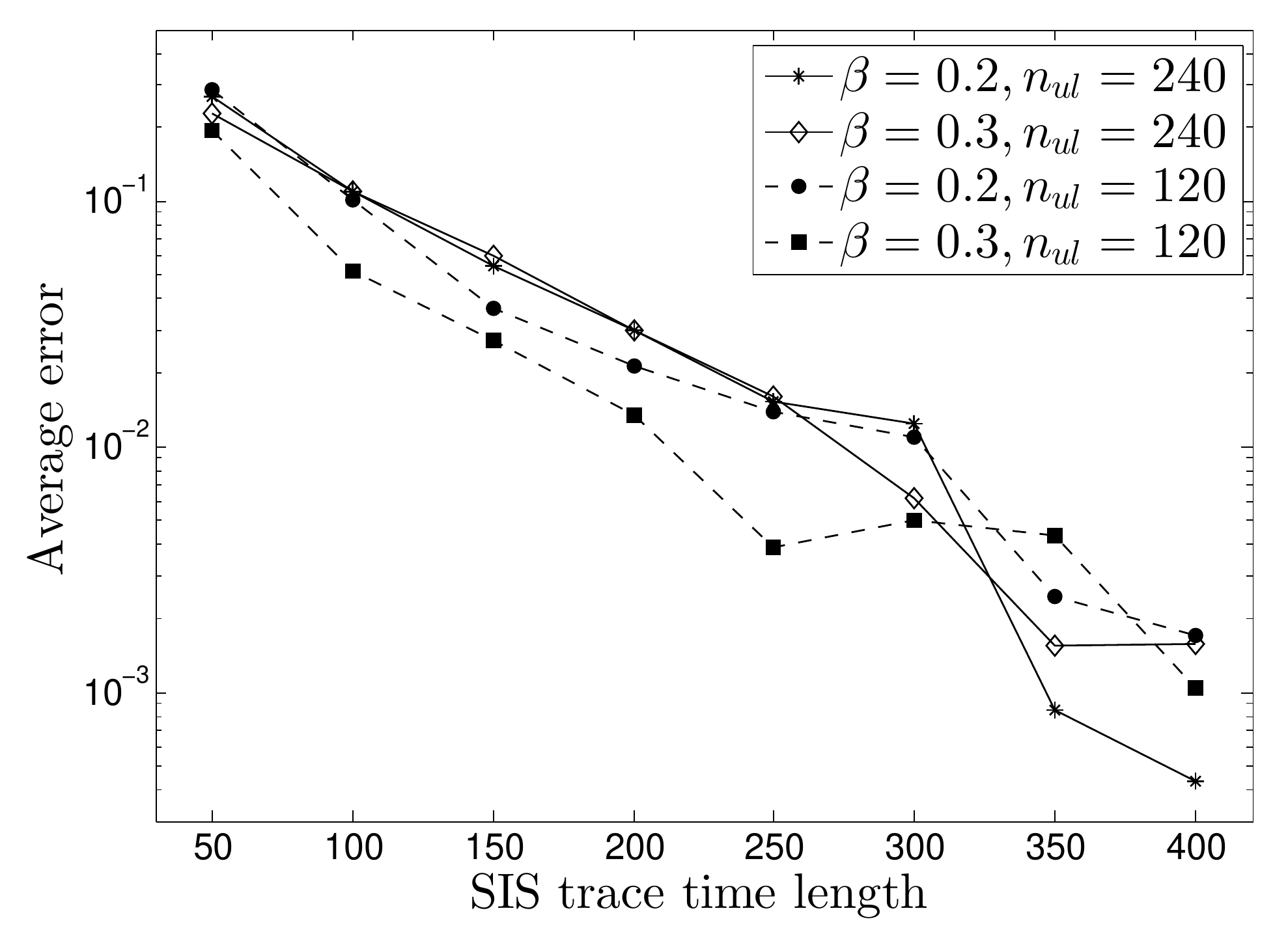}
    \caption{Result for the experiment in section \ref{exp1}.}
\label{av1}
\end{figure}

In this equation,  $\widehat{\mathrm{P}}_{i}$ is the posterior probability of the uncertain link $i$ and $n_{ul}$ the number of uncertain links. If the uncertain link is an actual link in the network,  $\gamma_{i}=1$, otherwise, $\gamma_{i}=0$.
We constructed two different sets of uncertain links with $120$ and $240$ links as explained before, and we performed the experiment independently on them. To demonstrate how the duration of SIS traces affects the inference, we obtained the posteriors using slices of the simulated SIS traces with increasing time length.  
Figure.(\ref{av1}) shows the average error defined in Eq.(\ref{er}) for different settings of the experiment. To demonstrate the accuracy of the inference, we can compare the average error of the prior distribution, which is 0.5, to the smallest values in the plot, which are on the order of $10^{-3}$. 
\subsection{Experiment using informative prior}\label{exp2}
\begin{figure}[t]
	\subfloat[]{
		\includegraphics[width=1\columnwidth]{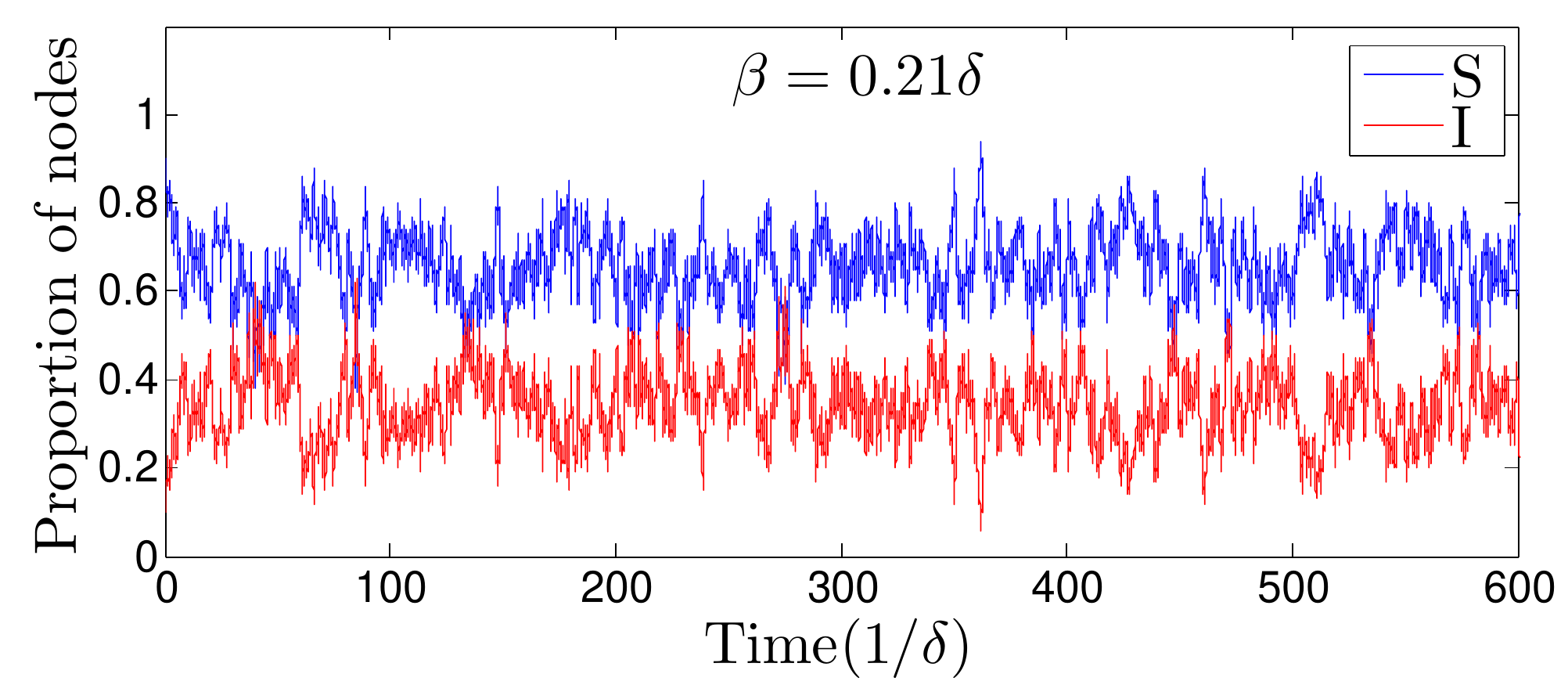}
\label{geosis}}%\\
	
	\subfloat[]{
		\includegraphics[width=1\columnwidth]{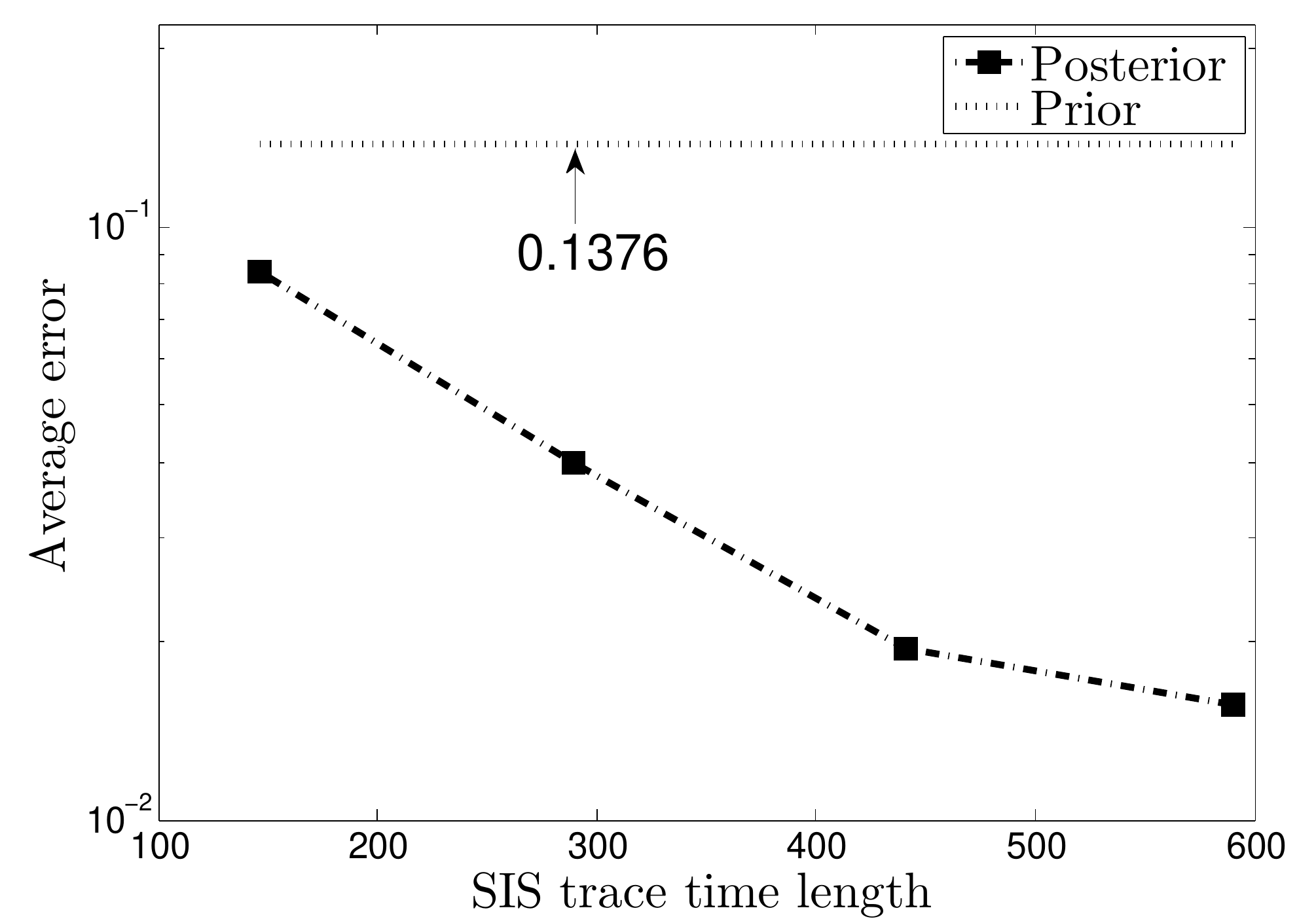}
\label{avgeo}}%
	
	\caption{ (a) Proportion of infected and susceptible individuals for the SIS trace used in the experiment  explained in section 
		\ref{exp2}, and (b)  result of the experiment that shows how the average error goes down with the increasing duration of the SIS trace} 
	\label{geo}%
\end{figure}

In the next experiment, we considered inferring all the links in a network of $100$ nodes using an SIS spreading trace. Here, we assumed the nodes are positioned on the vertices of a square grid, and we constructed a synthetic network by connecting any two nodes $a,b$ with a probability that decreases with their distance as
\begin{equation}
p(a,b)=\frac{0.3}{d(a,b)},
\end{equation}\label{geopr}
$d(a,b)$ is the Euclidean distance of nodes $a,b$ in a length unit defined by the grid's edges. This resulted in a network with $390$ edges. For this network, we assumed the transmission rates over the links are $\beta=0.21 \delta$ where $\delta$ is the recovery rate of the nodes. We simulated an SIS epidemic initiated from $10$ randomly infected nodes using an available epidemic simulator \cite{GEMFTool}. The population of infected and susceptible nodes through time is depicted in Fig.(\ref{geosis}). In this experiment, we assumed the set of uncertain links includes all the possible links between nodes. In other words, we assumed we do not know for a fact if any of the possible links exists in the network or not, and we tried to construct the network from the SIS spreading trace.
However, we had some prior information that we incorporated into the prior distribution of the uncertain link set in Eq.(\ref{prior1}) so that the factor terms become
\[\mathfrak{f}_{\lambda}(\lambda)=\frac{0.3}{d_{\lambda}}\ \delta(\lambda-\beta)+ \frac{0.7}{d_{\lambda}}\ \delta(\lambda),
\]    
where $d_{\lambda}$ is the distance between the two nodes adjacent to the link. In this experiment, we employed various segments of the simulated SIS trace all starting at $t=0$ but ending in different instants. 

\begin{figure}[t]
	\subfloat[]{
		\includegraphics[width=1\columnwidth]{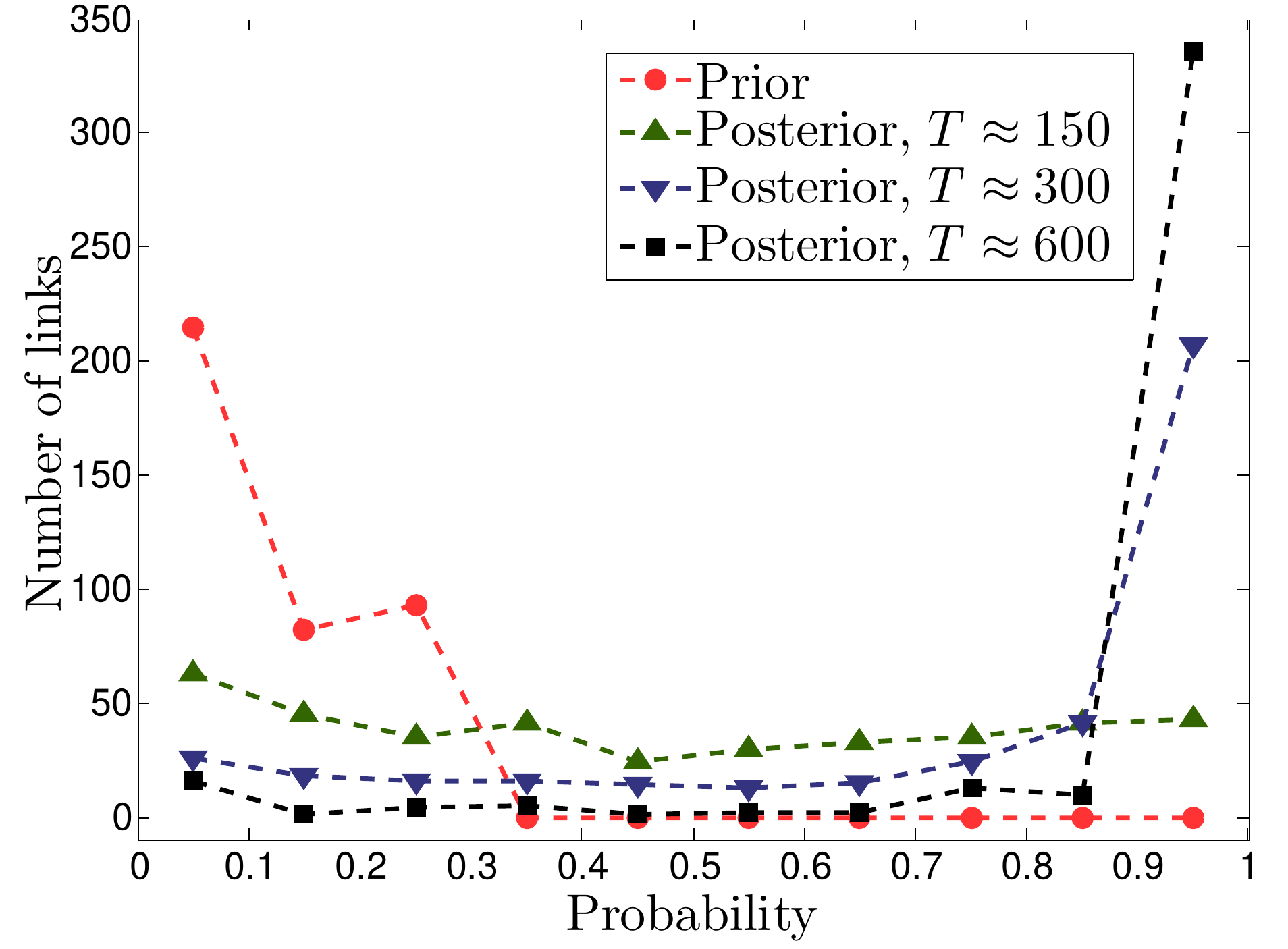}
		
\label{exist}}\\
	\subfloat[]{
		\includegraphics[width=1\columnwidth]{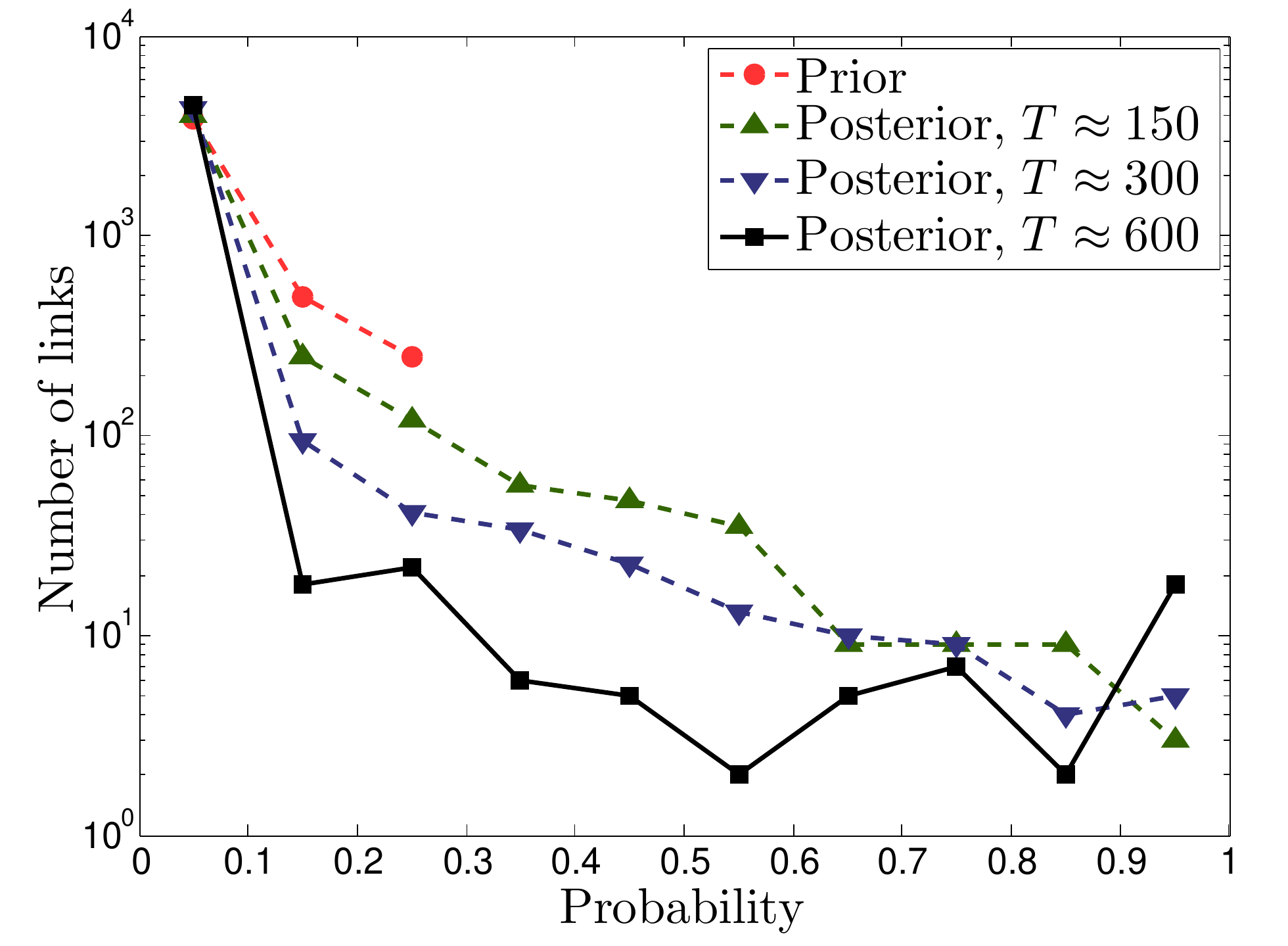}
\label{nonexist}
	}
	\caption{Result for the experiment in the section \ref{exp2}. The points in the plots show the number of links in  probability bins centered at $(0.5,1.5,\cdots,0.95)$ and with a width equal to $0.1$. Value of $T$ for each plot specifies the duration of SIS trace used for inferring the posterior probabilities. (a) Number of existing network links in different probability bins. (b)Number of non-existing links in each probability bin} 
	\label{hist}%
\end{figure}

Using each of these trace segments and the prior distribution in the equation above, we obtained the posterior probabilities for the links by performing Gibbs sampling. The average error of the posterior probabilities, defined in Eq.(\ref{er}), is shown in Fig(\ref{avgeo}). Furthermore, to show the accuracy of the inferred posterior we have plotted the histogram counts of the posterior probabilities for the actual links of the network in Fig.(\ref{exist}) and for the non-existing links in Fig.(\ref{nonexist}). As we explained, the uncertain link set includes all the possible $4950$ links between the nodes of the network. Among these uncertain links, $390$ are actual links of the network that generated the SIS trace, and the rest do not actually exist. However, the inference task provided a posterior probability for all the uncertain links. Fig.(\ref{exist}) shows the number of actual network links within different posterior probability bins as well as prior probabilities. We can see how the probabilities of the actual links move to larger values as the duration of SIS trace increases.  
On the other hand, Fig(\ref{nonexist}) shows a small number of non-existing links whose posterior probabilities are larger than $0.5$. However, the number of these links gets smaller as the time length of the SIS trace increases.

\section{CONCLUSION}
In this work, we investigated the inverse problem of stochastic spreading over a graph. Here, we proposed inferring the uncertain links of a network from the trace of continuous time SIS spreading. Due to the probabilistic nature of SIS spreading, it is not possible to determine the existence of an uncertain link with one hundred percent confidence, except in cases where an uncertain link is the sole mean of an infection event. Therefore, we formulated this inverse problem as a Bayesian inference problem, which provides a probability for the uncertain links. Furthermore, the Bayesian formulation of the problem allows us to incorporate other prior sources of information in the inference task. We provided a closed form formula to determine the probability of an uncertain link (Eq.\ref{marginal2}) when the calculations are tractable, and for other cases where we had a large number of uncertain links, we resorted to Gibbs sampling, which is a numerical method developed for Bayesian analysis. In our numerical experiments, we were able to assign the proper posterior probabilities to a large percentage of uncertain links. In other words, these probabilities were close to $1$ for almost all the actual network links and close to $0$ for the non-existing links.
\section*{Acknowledgement}
This material is based on work supported by the National Science Foundation under Grant No.CIF-1423411.

\bibliography{main}% Produces the bibliography via 
\end{document}